\documentclass{llncs}

\usepackage{bm,latexsym,amssymb,amsmath,mathrsfs,verbatim,stmaryrd}
\usepackage{fca,algorithm,algorithmic,url}

\title{A Survey on how Description Logic Ontologies Benefit from Formal Concept 
Analysis}
\author{Bar\i{}\c{s} Sertkaya}
\institute{SAP Research Center Dresden, Germany \\
\email{baris.sertkaya@sap.com}}

\renewcommand{\L}{{\mathcal L}}

\newcommand{\I}{\mathcal{I}}           
\newcommand{\dom}[1][\I]{\Delta^{#1}}  
\newcommand{\Int}[2][\I]{#2^{#1}}      

\newcommand{\T}{\mathcal{T}}                         
\newcommand{\A}{\mathcal{A}}                         

\newcommand{\Factpp}{FaCT++\xspace}            
\newcommand{\Racer}{\textsc{Racer}\xspace}     

\newcommand{\ALlang}[2]{\ensuremath{\mathcal{#1}#2}\xspace}
\newcommand{\FLE}{\ALlang{FLE}{}}
\newcommand{\EL}{\ALlang{EL}{}}
\newcommand{\ALE}{\ALlang{ALE}{}}
\newcommand{\ALC}{\ALlang{ALC}{}}

\begin{document}

\maketitle

\begin{abstract}
Although the notion of a concept as a collection of objects sharing certain
properties, and the notion of a conceptual hierarchy are fundamental to both
Formal Concept Analysis and Description Logics, the ways concepts are described 
and obtained differ significantly between these two research areas.  
Despite these differences, there have been several attempts to bridge
the gap between these two formalisms, and attempts to apply methods from one
field in the other. The present work aims to give an overview on the research
done in combining Description Logics and Formal Concept Analysis.
\end{abstract}

\section{Introduction}
\label{cha:introduction}

Formal Concept Analysis (FCA)~\cite{GaWi99}  is a field of applied mathematics 
that aims
to formalize the notions of a concept and a conceptual hierarchy by means
of mathematical tools.
On the other hand Description Logics (DLs)~\cite{BCNMP03}  are a class of 
logic-based knowledge 
representation formalisms that are used to represent the conceptual 
knowledge of an application domain in a structured way.
Although the notion of a concept as a collection of objects sharing certain
properties, and the notion of a conceptual hierarchy are fundamental to both 
FCA and DLs, the ways concepts are described and obtained differ significantly 
between these two research areas.  
In DLs, the relevant concepts of the application domain are formalized by 
so-called concept descriptions, which are expressions built from 
unary predicates (that are called atomic concepts), and binary predicates
(that are called atomic roles) with the help of the  concept constructors 
provided by the DL language. Then in a second step, these concept 
descriptions are used to describe 
properties of individuals occurring in the domain, and the roles are used to 
describe relations between these individuals.
On the other hand, in FCA, one starts with a so-called formal context, which 
in its simplest form is a way of specifying which attributes are satisfied by
which objects. 
A formal concept of such a context is a pair consisting of a 
set of objects called extent, and a set of attributes called intent such that 
the intent consists of exactly those attributes that the objects in the 
extent have in common, and the extent consists of exactly those objects that 
share all attributes in the intent.

There are several differences between these approaches. First, in FCA
one starts with a purely extensional description of the application domain, and 
then derives the formal concepts of this specific domain, which provide a 
useful structuring. 
In a way, in FCA the intensional knowledge is obtained from the extensional 
part of the knowledge.
On the other hand, in DLs the intensional definition of a concept is given 
independently of a specific domain (interpretation), and the description of 
the individuals is only partial. 
Second, in FCA the properties are atomic, and the intensional
description of a formal concept (by its intent) is just a conjunction
of such properties. DLs usually provide a richer language for the intensional 
definition of concepts, which can be seen as an expressive, yet decidable 
sublanguage of first-order predicate logic.

Despite these differences, there have been several attempts to bridge
the gap between these two formalisms, and attempts to apply methods from one 
field to the other. For example, 
there have been efforts to enrich FCA with more complex properties similar
to concept constructors in DLs
\cite{Zick91,PrSt99,Pred00,FeRS05,RHNV07}. 
On the other hand, 
DL research has benefited from
FCA methods to solve some problems encountered in knowledge representation
using DLs
\cite{Baad95,Stum96b,BaMo00,BaSe04,Rudo04,BaST04b,Rudo06,Sert06,BaST07,BGSS07,Sert07,Rudo08,BaDi08,BaDi09,BaSe09}.
%
The present work aims to give an overview on these works done for bridging the
gap between the two formalisms. 
In Section~\ref{sect:dls} we give a short introduction to DLs without going
into technical details. We assume that the reader is familiar with FCA. We
do not introduce FCA, we refer the reader to~\cite{GaWi99} for details.
 In Section~\ref{sect:existing_work} we summarize the 
existing work done by other researchers in the field. In 
Section~\ref{sect:contributions} we summarize our own contributions to
the field, and conclude with Section~\ref{sect:conclusion}.
\section{Description Logics}
\label{sect:dls}

Description Logics (DLs)~\cite{BCNMP03} are a class of knowledge representation 
formalisms that are used to represent the terminological knowledge of an 
application domain in a structured way. 
Since their introduction, DLs have been used 
in various application domains such as medical informatics, software 
engineering, configuration of technical systems, natural language processing, 
databases and web-based information systems. But their most notable success so 
far is the adoption of the DL-based language OWL\footnote{Web Ontology
Language. See \url{http://www.w3.org/TR/owl-features}}\cite{HoPH03} as the 
standard ontology language for the semantic web \cite{BLHL01}.

\subsubsection{Syntax}
In DLs, one formalizes the relevant notions of an application domain by 
\emph{concept descriptions}. A concept description is an expression built
from atomic concepts, which are unary predicates, and atomic roles, 
which are binary predicates, by using the concept constructors provided by the 
particular DL language in use. 
DL languages are identified with the concept constructors they allow. 
For instance the smallest propositionally closed language allowing for the 
constructors $\sqcap$ (conjunction), $\sqcup$ (disjunction), $\neg$ (negation), 
$\forall$ (value restriction) and $\exists$ (existential restriction) is called 
\ALC.

Typically, a \emph{DL knowledge base} consists of a \emph{terminological box}
(\emph{TBox}), which defines the terminology of an application domain, and an 
\emph{assertional box} (\emph{ABox}), which contains facts about a specific 
world.  In its simplest form, a TBox is a set of
\emph{concept definitions} of the form $A \equiv C$ that assigns the concept 
name $A$ to the concept description $C$. 
We call a finite set of \emph{general concept inclusion (GCI)} axioms a 
\emph{general TBox}. A GCI is an expression of the form $C \sqsubseteq D$, 
where $C$ and $D$ are two possibly complex concept descriptions. It states a 
subconcept/superconcept relationship between the two concept descriptions.
An ABox is a set of \emph{concept assertions} of the form $C(a)$, which means 
that the individual $a$ is an instance of the concept $C$, and 
\emph{role assertions} of the form $R(a,b)$, which means that the individual 
$a$ is in $R$-relation with individual $b$. 
%
%

For instance the following TBox contains the definition of a
landlocked country, which is a country that only has borders on land, and the
definition of an ocean country that has a border to an ocean.
\begin{displaymath}
\begin{array}{rcl}
\T := \{
\textsf{LandlockedCountry} & \equiv &
  \textsf{Country} \sqcap \forall \textsf{hasBorderTo}. \textsf{Land} \\
\textsf{OceanCountry} & \equiv &
\textsf{Country} \sqcap \exists \textsf{hasBorderTo}. \textsf{Ocean}
\}
\end{array}
\end{displaymath}
The following ABox states the facts about the individuals $Portugal$, 
$Austria$, and $Atlantic \ Ocean$.
\begin{displaymath}
\begin{array}{ccl}
\A & :=  & \{\textsf{LandlockedCountry}(Austria), \textsf{Country}(Portugal),
 \textsf{Ocean}(Atlantic \ Ocean), \\ 
& & \textsf{hasBorderTo}(Portugal,Atlantic \ Ocean)\}
\end{array}
\end{displaymath}
%

\subsubsection{Semantics}

The meaning of DL concepts is given by means of an 
\emph{interpretation} $\I$, which is a tuple consisting of a \emph{domain} $\dom$
and an \emph{interpretation function} $\Int{\cdot}$. The interpretation function
maps every concept occurring in the TBox to a subset of the domain, every
role to a binary relation on the domain, and every individual name occurring
in the ABox to an element of the domain. The meaning of complex concept 
descriptions is given inductively based on the constructors used in the
concept description. 

For instance, the concept description 
$\textsf{Country} \sqcap \exists \textsf{hasBorderTo}. \textsf{Ocean}$ is
interpreted as the intersection of the set of countries and the set of elements
of the domain that have a border to an ocean. 
We say that an interpretation 
$\I$ is a \emph{model} of a TBox $\T$ if it 
satisfies all concept definitions in $\T$, i.e., for every concept definition 
$A \equiv C$ in $\T$, it maps $A$ and $C$ to the same subset of the domain. 
Similarly, we say that $\I$ is a model of an 
ABox $\A$, if it satisfies
all concept and role assertions in $\A$, i.e., for every concept assertion
$A(a)$ in $\A$, the interpretation of $a$ is an element of the interpretation 
of $A$, and for every role assertion $r(a,b)$ the interpretation of $r$ 
contains the pair consisting of the interpretations of $a$ and $b$. 
The 
semantics of DL ABoxes is the \emph{open-world semantics}, i.e., absence of
information about an individual is not interpreted as negative information,
but it only indicates lack of knowledge about that individual.

\subsubsection{Inferences}
In an application, once we get a description of the application domain using 
DLs as described above, we can make inferences, i.e., deduce implicit 
consequences from the 
explicitly represented knowledge. The basic inference on concept descriptions
is \emph{subsumption}. Given two concept descriptions $C$ and $D$, the
subsumption problem $C \sqsubseteq D$ is the problem of checking whether 
the concept description $D$ is more general than the concept description 
$C$. In other words, it is the problem of determining whether the first 
concept always, i.e., in every interpretation denotes a subset of the set 
denoted by the second one. We say
that $C$ is subsumed by $D$ w.r.t. a TBox $\T$, if in every model
of $\T$, $D$ is more general than $C$, i.e., the interpretation of $C$ is a 
subset of the interpretation of $D$. We denote this as $C \sqsubseteq_\T D$. 
For instance, in the example above, the concepts 
$\textsf{LandlockedCountry}$ and $\textsf{OceanCountry}$ are both trivially 
subsumed by the concept $\textsf{Country}$. 

The typical inference problem for ABoxes is 
\emph{instance checking}, which is the problem of deciding 
whether the interpretation of a given individual is an element of the 
interpretation of a given concept in every common model of the TBox and
the ABox. For instance, from $\T$ and $\A$ given above  it follows 
that $Portugal$ is an ocean country, although $\A$ does not 
contain the assertion $\textsf{OceanCountry}(Portugal)$.
Modern \emph{DL systems} like \Factpp~\cite{TsHo06}, \Racer~\cite{HaMo01b}, 
Pellet~\cite{SiPa04}, KAON2~\cite{Moti06}, Hermit~\cite{MoSH09} and 
CEL~\cite{BaLS06}
provide their users with inference services that solve these inference 
problems, which are also known as \emph{standard inferences}. 

\section{Existing work on DLs and FCA}
\label{sect:existing_work}

The existing work done by other researchers
towards bridging the gap between FCA und DLs, and attempts to apply 
methods from one field to the other can roughly be collected under two 
categories:

\begin{itemize}
 \item efforts to enrich the language of FCA by borrowing constructors from
 DL languages \cite{Zick91,PrSt99,Pred00,FeRS05,RHNV07}
 \item efforts to employ FCA methods in the solution of problems encountered in
 knowledge representation with DLs 
 \cite{Baad95,Stum96b,BaMo00,Rudo04,Rudo06,Rudo08,BaDi08,CTND08,CTND08b,BaDi09}
\end{itemize}
Below we are going to discuss some of these efforts briefly.



\subsection{Enriching FCA with DL constructors}

\subsubsection{Theory-driven logical scaling}

In \cite{PrSt99}, Prediger and Stumme have used DLs in \emph{Conceptual
Information Systems}, which are data analysis tools based on FCA. They can be
used to extract data from a relational database and to store it in a formal 
context by using so-called \emph{conceptual scales}. Prediger and Stumme have 
combined DLs with attribute exploration in order to define a new kind of
conceptual scale. In this approach, DLs provide a rich language to specify 
which FCA attributes cannot occur together, and a DL reasoner is used 
during the attribute exploration process as an expert to answer the
implication questions, and to provide a counterexample whenever the implication
does not hold.

\subsubsection{Terminological attribute logic}

In \cite{Pred00}, Prediger has worked on introducing logical constructors 
into FCA. She has enriched FCA with relations, existential and universal
quantifiers, and negation, obtaining a language like the DL \ALC, which
she has called \emph{terminologische Merkmalslogik} (\emph{terminological
attribute logic}\footnote{This translation is ours.}).
In the same work she has also presented applications of her approach in 
enriching formal contexts with new knowledge, applications in many valued
formal contexts, and applications for so-called \emph{scales}, which are
formal contexts that are used to obtain a standard formal context from a 
many valued formal context.

\subsubsection{Relational concept analysis}

In \cite{RHNV07}, Rouane et al. have presented a combination of FCA and DLs 
that is called \emph{relational concept analysis}. It is an adaptation of 
FCA that is intended for analyzing objects described by relational attributes in
data mining. The approach is based on a collection of formal
contexts called \emph{relational context family} and relations between these
contexts. The relations between the contexts are binary relations between
pairs of object sets that belong to two different contexts. Processing these
contexts and relations with relational concept analysis methods yields a set
of concept lattices (one for each input context) such that the formal 
concepts in different lattices are linked by relational attributes, which are
similar to roles in DLs, or associations in UML. One distinguishing feature of
this approach from the other efforts that introduce relations into FCA is that 
the formal concepts and relations between formal concepts of different
contexts can be mapped into concept descriptions in a sublanguage of \ALE, 
which is called $\mathcal{FL}^-\mathcal{E}$ in \cite{RHNV07}.
$\mathcal{FL}^-\mathcal{E}$ allows for conjunction, value restriction, 
existential restriction, and top and bottom concepts. In this approach, after
the formal concepts and relations have been obtained and mapped into 
$\mathcal{FL}^-\mathcal{E}$ concept descriptions, DL reasoning is used to
classify and check the consistency of these descriptions.

\subsection{Applying FCA methods in DLs}

\subsubsection{Subsumption hierarchy of conjunctions of DL concepts}

In \cite{Baad95}, Baader has used FCA for an efficient computation of an 
extended subsumption hierarchy of a set of DL concepts. More precisely, he used 
attribute exploration for computing the subsumption hierarchy of all 
conjunctions of a set of DL concepts. The main motivation for this work was
to determine the interaction between defined concepts, which might not easily
be seen by just looking at the subsumption hierarchy of defined concepts.
In order to explain this, the following example has been given: 
assume that the defined concept \textsf{NoDaughter} stands for those people who
have no daughters, \textsf{NoSon} stands for those people who have no sons,
and \textsf{NoSmallChild} stands for those people who have no small children.
Obviously, there is no subsumption relationship between these three concepts.
On the other hand, the conjunction $\textsf{NoDaughter} \sqcap
\textsf{NoSon}$ is subsumed by \textsf{NoSmallChild}, i.e., if an individual
$a$ belongs to \textsf{NoSon} and \textsf{NoDaughter}, it also belongs to
\textsf{NoSmallChild}. However, this cannot be derived from the information
that $a$ belongs to \textsf{NoSon} and \textsf{NoDaughter} by just looking
at the subsumption hierarchy. This small example demonstrates that runtime
inferences concerning individuals can be made faster by precomputing the
subsumption hierarchy not only for defined concepts, but also for all
conjunctions of defined concepts.

To this purpose, Baader defined a formal context whose attributes were
the defined DL concepts, and whose objects were all possible counterexamples to 
subsumption relationships, i.e., interpretations together with an element of
the interpretation domain. This formal context has the property that its 
concept lattice is isomorphic to the required subsumption hierarchy, namely
the subsumption hierarchy of conjunctions of the defined DL concepts. 
However, this formal context has the disadvantage that a standard subsumption 
algorithm can not be used as expert for this context within attribute 
exploration. In order to overcome 
this problem, the approach was reconsidered in \cite{BaSe04} and a new 
formal context that has the same properties but for which a usual subsumption
algorithm could be used as expert was introduced. 

\subsubsection{Subsumption hierarchy of conjunctions and disjunctions of DL
concepts}

In \cite{Stum96b}, Stumme has extended the abovementioned subsumption hierarchy
further with disjunctions of DL concepts. More precisely, he 
presented how the complete lattice of all possible combinations of conjunctions
and disjunctions of the concepts in a DL TBox can be computed
by using FCA. To this aim, he used another knowledge acquisition tool of
FCA instead of attribute exploration, namely \emph{distributive concept 
exploration} \cite{Stum98}. In the lattice computed by this method, the 
supremum of two DL concepts in the lattice corresponds to the disjunction of 
these concepts.

\subsubsection{Subsumption hierarchy of least common subsumers}

In \cite{BaMo00} Baader and Molitor have used FCA for supporting bottom-up
construction of DL knowledge bases. In the bottom-up approach, the knowledge
engineer does not directly define the concepts of her application domain, but 
she gives typical examples of a concept, and the system comes up with a 
concept description for these examples. The process of computing such a 
concept description consists of first computing the \emph{most specific
concepts} that the given examples belong to, and then computing the
\emph{least common subsumer} of these concepts.
Here the choice of examples is 
crucial for the quality of the resulting concept description. If the examples
are too similar, the resulting concept description will be too specific;
conversely, if they are too distinct, the resulting concept description will
be too general. In order to overcome this, Baader and Molitor have used 
attribute exploration for computing the subsumption hierarchy of all least
common subsumers of a given set of concepts. In this hierarchy one can easily
see the position of the least concept description that the chosen examples
belong to, and decide whether these examples are appropriate for obtaining
the intended concept description. However, there may be exponentially many 
least common subsumers, and depending on the DL in use, both the least common 
subsumer computation and subsumption test can be expensive operations. 
The use of attribute exploration provides us with complete information on how 
this hierarchy looks like without explicitly computing all least common 
subsumers and classifying them. 

\subsubsection{Relational exploration}

In his Ph.D thesis \cite{Rudo06}, Rudolph has combined DLs and FCA for 
acquiring complete relational knowledge about an application domain. In
his approach, which he calls \emph{relational exploration}, he uses DLs
for defining FCA attributes, and FCA for refining DL knowledge bases. More
precisely, DLs makes use of the interactive knowledge 
acquisition method of FCA, and FCA benefits from DLs in terms of expressing
relational knowledge.

In \cite{Rudo04,Rudo06}, Rudolph uses the DL \FLE for this purpose, which is
the DL that allows for the constructors 
conjunction, existential restriction, and value restriction. In his previous 
work \cite{Rudo03}, he uses the DL \EL, which allows for the constructors
conjunction and existential restriction. In both cases,
he defines the semantics by means of a special pair of formal contexts
called \emph{binary power context family}, which are used for expressing 
relations in FCA. Binary power context families have also been used for giving
semantics to \emph{conceptual graphs}. In order to collect information about
the formulae expressible in \FLE, in \cite{Rudo04,Rudo06} he defines a formal 
context called \FLE-\emph{context}. The attributes of this formal context
are \FLE-concept descriptions, and the objects are the elements of the 
domain over which these concept descriptions are interpreted. In this context,
an object $g$ is in relation with an attribute $m$ if and only if $g$ is in
the interpretation of $m$. Thus, an implication holds in this formal context
if and only if in the given model the concept description resulting from the conjunction 
of the attributes in the premise of the implication is subsumed by the 
concept description formed from the conclusion. This is how implications in
\FLE-contexts give rise to subsumption relationships between \FLE concept
descriptions. 

In order to obtain \emph{complete} knowledge about the subsumption relationships
in the given model between arbitrary \FLE concepts, Rudolph gives a multi-step exploration 
algorithm. In the first step of the algorithm, he starts with an \FLE-context
whose attributes are the atomic concepts occurring in a knowledge base. In
exploration step $i+1$, he defines the set of attributes as the union of the
set of attributes from the first step and the set of concept descriptions 
formed by universally quantifying all attributes of the context at step $i$ 
w.r.t.\,  all atomic roles, and the set of concept descriptions formed by 
existentially quantifying all concept intents of the context at step $i$ w.r.t 
all atomic roles.
Rudolph points out that, at an exploration step, there can be some concept
descriptions in the attribute set that are equivalent, i.e., attributes that 
can be reduced. To this aim, he introduces a method that he calls 
\emph{empiric attribute reduction}. In principle, it is possible to carry out 
infinitely many exploration steps, which means that the algorithm
will not terminate. In order to guarantee termination, Rudolph restricts the 
number of exploration steps. After carrying out $i$ steps of exploration, it is 
then possible to 
decide subsumption (w.r.t. the given model) between any \FLE concept descriptions up to 
role depth $i$ just by using the implication bases obtained as a result of
the exploration steps. In addition, he also characterizes the cases where finitely many 
steps are sufficient to acquire complete information for deciding subsumption between 
\FLE concept descriptions with arbitrary role depth. Rudolph argues that his method can 
be used to support the knowledge engineers in designing, building and refining 
DL ontologies. This method has been implemented in the tool 
Relexo.\footnote{\url{http://relexo.ontoware.org}}

\subsubsection{Exploring Finite Models in the DL $\EL_{gfp}$}

In~\cite{BaDi08} Baader and Distel have extended classical FCA in order to
provide support for analyzing relational structures by using efficient FCA
algorithms. In this approach the atomic attributes are replaced by complex
formulae in some logical language, and data is represented using relational
structures rather than just formal contexts. This extension is later 
instantiated with atrributes defined in the DL \EL, and with relational 
structures defined over a signature of unary and binary predicates, i.e., models
for \EL. In this setting an implication corresponds to a GCI in \EL. This 
approach at the first sight seems to be very close to the
approach introduced in~\cite{Rudo04,Rudo06}. One of the main differences
between these approaches is that in~\cite{BaDi08} the authors
use one context with infinitely many complex attributes, whereas 
in~\cite{Rudo06} Rudolph uses an infinite family of contexts, each having
finitely many attributes that are obtained by restricting the role depth
of concepts. In~\cite{BaDi08} the authors additionally show that for the
DLs \EL and $\EL_{gfp}$, which extends  \EL with cyclic concept definitions
interpreted with \emph{greatest fixpoint} semantics, 
the set of GCIs holding in a finite model always has a finite basis. That is,
there is always a finite subset of the infinitely many GCIs from which the
rest follows.
Later in~\cite{BaDi09} the authors have shown how to compute this basis 
efficiently by using methods from FCA. In a follow-up paper~\cite{Dist10b}, 
Distel has described how this method can be modified to allow ABox individuals 
as counterexamples to GCIs.

\section{Contributions to combining DLs and FCA}
\label{sect:contributions}

Our contribution to the DL research by means of FCA methods falls
mainly under two topics: 1) supporting bottom-up construction of DL knowledge
bases, 2) completing DL knowledge bases. In Section 
\ref{subsec:supportingBottomup} we briefly describe the use of FCA 
in the former,
and in Section \ref{subsect:completing} we briefly describe the use of FCA 
in the latter contribution.

\subsection{Supporting bottom-up construction of DL Ontologies}
\label{subsec:supportingBottomup}

Traditionally, DL knowledge bases are built in a top-down manner, in the 
sense that first the relevant notions of the domain are formalized by 
concept descriptions, and then these concept descriptions are used to 
specify properties of the individuals occurring in the domain. However, this 
top-down approach is not always adequate. On the one hand, it might not 
always be intuitive which notions of the domain are the relevant ones for
a particular application. On the other hand, even if this is intuitive, it might
not always be easy to come up with a clear formal description of these notions, 
especially for a domain expert who is not an expert in knowledge engineering.
In order to overcome this, in \cite{BaKM99} a new approach, called
``bottom-up approach''\index{bottom-up approach}, was introduced for 
constructing DL knowledge bases. In this approach,
instead of directly defining a new concept, the domain expert introduces
several typical examples as objects, which are then automatically generalized
into a concept description by the system. This description is then offered to 
the domain expert as a possible candidate for a definition of the concept.
The task of computing such a concept description can be split into two 
subtasks:
\begin{itemize}
\item computing the most specific concepts of the given objects, 
\item and then computing the least common subsumer of these concepts.  
\end{itemize}
The \emph{most specific concept}\index{most specific concept} (msc) of an 
object $o$  is the most 
specific concept description $C$ expressible in the given DL language that has 
$o$ as an instance. 
The \emph{least common subsumer (lcs)}\index{least common subsumer}
of concept descriptions 
$C_1,\ldots, C_n$ is the most specific concept description $C$ expressible in 
the given DL language that subsumes $C_1,\ldots, C_n$.
The problem of computing the lcs and (to a more limited extent) the msc has 
already been investigated in the literature 
\cite{BaKM99,KuMo01,Baad03b}.

The methods for computing the least common subsumer are restricted
to rather inexpressive descriptions logics not allowing for disjunction
(and thus not allowing for full negation).
In fact, for languages with disjunction, the lcs of a collection of concepts
is just their disjunction, and nothing new can be learned from building
it. In contrast, for languages without disjunction, the lcs extracts
the ``commonalities'' of the given collection of concepts.
Modern DL systems like \Factpp~\cite{Horr98,TsHo06}, 
\Racer~\cite{HaMo01b}, Pellet~\cite{SiPa04}, and Hermit~\cite{MoSH09}  are 
based on very expressive DLs, and there exist large knowledge
bases that use this expressive power and can be processed
by these systems. 
In order to allow the user
to re-use concepts defined in such existing knowledge bases and still
support the user in defining new concepts with the bottom-up
approach sketched above, in \cite{BaST04,BaST04b,BaST07} 
we have proposed the following \emph{extended bottom-up approach}:
assume that there is a fixed 
\emph{background terminology} defined in an expressive DL; e.g., a large
ontology written by experts, which the user has bought from some ontology
provider. The user then wants to extend this terminology in order
to adapt it to the needs of a particular application domain. However,
since the user is not a DL expert, he employs a less expressive DL and
needs support through the bottom-up approach when building this user-specific 
extension of the background terminology. There are several reasons for the
user to employ a restricted DL in this setting: first, such a restricted DL 
may be easier to comprehend and use for a non-expert; second, it may allow for 
a more intuitive graphical or frame-like user interface; third, to use the 
bottom-up approach, the lcs must exist and make sense, and it must be possible 
to compute it with reasonable effort.

To make this more precise, consider a background terminology (TBox) $\T$ 
defined in an expressive DL $\L_2$. When defining new concepts, the user 
employs only a sublanguage $\L_1$ of $\L_2$, for which computing the lcs makes 
sense. However, in addition to primitive concepts and roles, the concept 
descriptions written in the DL $\L_1$ may also contain names of concepts 
defined in $\T$.  Let us call such concept descriptions $\L_1(\T)$-concept
descriptions. Given $\L_1(\T)$-concept descriptions $C_1,\ldots, C_n$,
we want to compute their lcs in $\L_1(\T)$, i.e., the least
$\L_1(\T)$-concept description that subsumes $C_1,\ldots, C_n$ w.r.t.\ $\T$.
In \cite{BaST04b,BaST07} we have considered the case where 
$\L_1$ is the DL $\ALE$ and $\L_2$ is the DL $\ALC$, and shown the 
following result:
\begin{itemize}
\item
  If $\T$ is an acyclic $\ALC$-TBox, then the lcs w.r.t.\ $\T$ of
  $\ALE(\T)$-concept descriptions always exists.
\end{itemize}
%
Unfortunately, the proof of this result does not yield a practical
algorithm. Due to this, in \cite{BaST04b,BaST07,Sert07} we have developed a 
more practical 
approach. Assume that $\L_1$ is a DL for which least common subsumers
(without background TBox) always exist.
Given $\L_1(\T)$-concept descriptions $C_1, \ldots, C_n$, one can compute a 
common subsumer w.r.t.\ $\T$ by just ignoring $\T$, i.e., by treating the 
defined names in $C_1, \ldots, C_n$ as primitive and computing the lcs of
$C_1, \ldots, C_n$ in $\L_1$. However, the common subsumer obtained this way 
will usually be too general. In \cite{BaST04b,BaST07,Sert07}, 
work we presented a method for computing ``good'' common subsumers w.r.t.\ 
background TBoxes, which may not be the \emph{least} common subsumers, but 
which are better than the common subsumers computed by ignoring the TBox.
In the present work we do not give the gcs algorithm in detail. We only 
demonstrate it on an example. The algorithm is described in detail 
in~\cite{BaST07}.
\begin{example}
As a simple example, consider the $\ALC$-TBox $\T$:
$$
\begin{array}{rcl}
\textsf{NoSon} &\equiv& \forall\textsf{has-child}.\textsf{Female}, \\
\textsf{NoDaughter} &\equiv& \forall\textsf{has-child}.\neg\textsf{Female},\\
\textsf{SonRichDoctor} &\equiv&
  \forall\textsf{has-child}.(\textsf{Female}\sqcup (\textsf{Doctor}\sqcap\textsf{Rich})),\\
\textsf{DaughterHappyDoctor} &\equiv&
  \forall\textsf{has-child}.(\neg\textsf{Female}\sqcup (\textsf{Doctor}\sqcap\textsf{Happy})),\\
\textsf{ChildrenDoctor} &\equiv& \forall\textsf{has-child}.\textsf{Doctor},
\end{array}
$$
and the $\ALE$-concept descriptions
$$
\begin{array}{rcl}
C &:=& \exists\textsf{has-child}.(\textsf{NoSon}\sqcap\textsf{DaughterHappyDoctor}),\\
D &:=& \exists\textsf{has-child}.(\textsf{NoDaughter}\sqcap\textsf{SonRichDoctor}).
\end{array}
$$
By ignoring the TBox, we obtain the $\ALE(\T)$-concept description
$\exists\textsf{has-child}.\top$
as a common subsumer of $C, D$. However, if
we take into account that both 
$\textsf{NoSon}\sqcap\textsf{DaughterHappyDoctor}$ and
$\textsf{NoDaughter}\sqcap\textsf{SonRichDoctor}$ are subsumed by
the concept $\textsf{ChildrenDoctor}$, then we
obtain the more specific common
subsumer  $\exists\textsf{has-child}.\textsf{ChildrenDoctor}$.
The gcs of $C, D$ is even more specific. In fact,
the least conjunction of (negated) concept names subsuming both
$\textsf{NoSon}\sqcap\textsf{DaughterHappyDoctor}$ and
$\textsf{NoDaughter}\sqcap\textsf{SonRichDoctor}$ is
$$
\textsf{ChildrenDoctor}\sqcap\textsf{DaughterHappyDoctor}\sqcap\textsf{SonRichDoctor},
$$
and thus the gcs of $C, D$ is
$$
\exists\textsf{has-child}.(\textsf{ChildrenDoctor}\sqcap\textsf{DaughterHappyDoctor}\sqcap\textsf{SonRichDoctor}).
$$
The conjunct $\textsf{ChildrenDoctor}$ is actually redundant since it is implied
by the remainder of the conjunction.
\hfill $\diamond$
\end{example}

In order to implement the gcs algorithm, we must be able to compute
the smallest conjunction of (negated) concept names that subsumes two such 
conjunctions $C_1$ and $C_2$ w.r.t.\ $\T$. In principle, one can compute this 
smallest conjunction by testing, for every (negated) concept name whether it 
subsumes both $C_1$ and $C_2$ w.r.t.\ $\T$, and then take the conjunction of 
those (negated) concept names for which the test was positive.
However, this results in a large number of (possibly expensive) calls to
the subsumption algorithm for $\L_2$ w.r.t.\ (general or (a)cyclic) TBoxes.
Since, in our application scenario (bottom-up construction of DL knowledge bases
w.r.t.\ a given background terminology), the TBox $\T$ is assumed to be
fixed, it makes sense to precompute this information.

This is where FCA comes into play. By using
the attribute exploration method \cite{Gant84} (possibly with
background knowledge \cite{Gant99,GaKr99,GaKr05}), we compute the 
abovementioned smallest conjunction, which is required for computing a gcs.
To this purpose
we define a formal context whose concept lattice is isomorphic
to the subsumption hierarchy we are interested in. In general, the subsumption 
relation induces a partial order, and not a lattice structure on concepts. 
However, in the case of conjunctions of (negated) concept names, all infima 
exist, and thus also all suprema, i.e., this hierarchy is a complete lattice.
The experimental results in~\cite{BaST07} have shown that the use of 
this hierarchy and its use in gcs computation are indeed quite efficient.

%

\subsection{Completing DL Ontologies}
\label{subsect:completing}

The standardization of OWL~\cite{HoPH03} as the ontology language for the
semantic web~\cite{BLHL01} led to the fact that 
several ontology editors 
like Prot\'eg\'e~\cite{KFNM04}, and Swoop~\cite{KPSG+06a} now support OWL, and 
ontologies written in OWL
are employed in more and more applications.  As the size of 
these ontologies grows, tools that support improving their quality 
become more important. The tools available until now use DL reasoning
to detect inconsistencies and to infer consequences, i.e., implicit
knowledge that can be deduced from the explicitly represented
knowledge. There are also promising approaches that allow to pinpoint
the reasons for inconsistencies and for certain consequences, and that
help the ontology engineer to resolve inconsistencies and to remove
unwanted consequences \cite{ScCo03,KPSG06b,KPHS07,HoPS08,BaPe10,PeSe10}. These 
approaches
address the quality dimension of \emph{soundness} of an ontology, both
within itself (consistency) and w.r.t.\ the intended application
domain (no unwanted consequences). In 
\cite{BGSS07b,BGSS07} we have considered a different quality 
dimension: \emph{completeness}. We have
provided a basis for formally well-founded techniques and tools that
support the ontology engineer in checking whether an ontology contains
all the relevant information about the application domain, and to
extend the ontology appropriately if this is not the case.

As already mentioned, a DL knowledge base (nowadays 
often called ontology) usually consists of two parts, the terminological part 
(TBox), which defines concepts and also states additional constraints 
(GCIs) on the interpretation 
of these concepts, and the assertional part (ABox), which describes individuals 
and their relationship to each other and to concepts.  Given an application
domain and a DL knowledge base describing it, we can ask whether
the knowledge base contains all the relevant information
about the domain:
\begin{itemize}
\item
  Are all the relevant constraints that hold between concepts in the domain 
  captured by the TBox?
\item
  Are all the relevant individuals existing in the domain represented in the 
  ABox?
\end{itemize}

As an example, consider the OWL ontology for human protein phosphatases that has
been described and used in \cite{WBHL+05}. This ontology was developed
based on information from peer-reviewed publications.
The human protein phosphatase family has been well characterised
experimentally, and detailed knowledge about different classes of such
proteins is available. This knowledge is represented in the terminological part
of the ontology. Moreover, a large set of human
phosphatases has been identified and documented by expert biologists.
These are described as individuals in the assertional part of the ontology.
One can now ask whether the information about protein phosphatases contained
in this ontology is complete: are all the relationships that hold among
the introduced classes of phosphatases captured by the constraints
in the TBox, or are there relationships that hold in the domain, but do not
follow from the TBox? Are all possible kinds of human protein phosphatases 
represented by individuals in the ABox, or are there phosphatases that have 
not yet been included in the ontology or even not yet have been identified?

Such questions cannot be answered by an automated tool alone. Clearly,
to check whether a given relationship between concepts---which does
not follow from the TBox---holds in the domain, one needs to ask a
domain expert, and the same is true for questions regarding the
existence of individuals not described in the ABox.  The r\^ole of the
automated tool is to ensure that the expert is asked as few questions
as possible; in particular, she should not be asked trivial questions,
i.e., questions that could actually be answered based on the
represented knowledge.  In the above example, answering a non-trivial
question regarding human protein phosphatases may require the
biologist to study the relevant literature, query existing protein
databases, or even to carry out new experiments.  
Thus, the expert may be prompted to acquire new biological knowledge.

The attribute exploration method 
of FCA
has proved to be a successful knowledge acquisition method in various
application domains. One of the earliest
applications of this approach is described in \cite{Will82}, where the
domain is lattice theory, and the goal of the exploration process is
to find, on the one hand, all valid relationships between
properties of lattices (like being distributive), and, on the other
hand, to find counterexamples to all the relationships that do not
hold. To answer a query whether a certain relationship holds, the
lattice theory expert must either confirm the relationship (by using
results from the literature or by carrying out a new proof for this fact),
or give a counterexample (again, by either finding one in the
literature or constructing a new one).

Although this sounds very similar to what is needed in our case, we
cannot directly use this approach. The main reason is the open-world
semantics of description logic knowledge bases. Consider an individual
$i$ from an ABox $\A$ and a concept $C$ occurring in a TBox $\T$. If we
cannot deduce from the TBox $\T$ and $\A$ that $i$ is an instance of $C$,
then we do not assume that $i$ does not belong to $C$. Instead, we
only accept this as a consequence if $\T$ and $\A$ imply that $i$ is an
instance of $\neg C$. Thus, our knowledge about the relationships
between individuals and concepts is incomplete: if $\T$ and $\A$ imply
neither $C(i)$ nor $\neg C(i)$, then we do not know the relationship
between $i$ and $C$. In contrast, classical FCA and attribute
exploration assume that the knowledge about objects is complete:
a cross in row $g$ and column $m$ of a formal context says that object $g$ has 
attribute $m$, and the absence of a cross is interpreted as saying that 
$g$ does not have $m$.

There has been some work on how to extend FCA and attribute
exploration from complete knowledge to the case of partial knowledge
\cite{Gant99,BuHo00,Holz04a,Holz04b,BuHo05,Rudo06}, and how to evaluate 
formulas in formal contexts that do not contain complete information 
\cite{Obie02}. However, these works are based on assumptions that are different 
from ours. In particular, they 
assume that the expert cannot answer all queries and, as a consequence, the 
knowledge obtained after the exploration process may still be incomplete and 
the relationships between concepts that are produced in the end fall into
two categories: relationships that are valid no matter how the
incomplete part of the knowledge is completed, and relationships that
are valid only in some completions of the incomplete part of the
knowledge. In contrast, our intention is to complete the knowledge base, i.e., 
in the end we want to have complete knowledge about these
relationships. What may be incomplete is the description of
individuals used during the exploration process.

In~\cite{BGSS07,Sert07} we have introduced an extension of FCA that can 
deal with partial knowledge.
This extension is based on the notion of a \emph{partial context} that consists 
of a set of \emph{partial object descriptions} (\emph{pod}). A pod is a tuple 
$(A,S)$ where $A$ represents the set of attributes that the pod is known to 
have, and $S$ represents the set of attributes that the pod is known not to 
have. $A$ and $S$ are disjoint and their union need not be the whole attribute 
set, i.e., for some attributes it might be unknown whether the pod has this 
attribute or not. We say that a pod $(A,S)$ \emph{refutes} an implication
$L \rightarrow R$ if $L \subseteq A$ and $R \cap S \neq \emptyset$. We also
say that a partial context refutes an implication if there is a pod in this 
partial context that refutes this implication. Based on these, we define the
notion of an \emph{undecided implication}, which is an implication that
does not follow from a given set of implications, and that is not refuted by
a partial context. Then the attribute exploration method for partial contexts
can be formulated as enumerating undecided implications as efficient as
possible. In~\cite{BGSS07,Sert07} we have described a version of attribute
exploration algorithm that works for this setting, and  proved that this
algorithm terminates and it is correct. Later we have shown
that given a DL knowledge base $(\T,\A)$, any individual in $\A$ gives
rise to a pod, and thus $\A$ induces a partial context. This enables us
to use our attribute exploration algorithm on partial contexts for 
finding completing DL knowledge bases. As a result of running this algorithm
on a DL knowledge base, the knowledge base is complete w.r.t. an intended
interpreation, i.e., if an implication holds in this interpreation then 
it also follows from the TBox, and if not then the ABox contains a 
counterexample to this implication. For details of the attribute exploration
on partial contexts and its application to DL ontologies we refer the reader
to~\cite{BGSS07,Sert07}, and demonstrate on a small example how it works.
\begin{example}

Let our TBox $\T_{countries}$ contain the following concept definitions:
\begin{displaymath}
\begin{array}{rcl}
\textsf{AsianCountry} & \equiv & \textsf{Country} \sqcap \exists \textsf{hasTerritoryIn}.\{Asia\} \\
\textsf{EUmember} & \equiv & \textsf{Country} \sqcap \exists \textsf{memberOf}.\{EU\} \\
\textsf{EuropeanCountry} & \equiv & \textsf{Country} \sqcap \exists \textsf{hasTerritoryIn}.\{Europe\} \\
\textsf{G8member} & \equiv & \textsf{Country} \sqcap \exists \textsf{memberOf}.\{G8\} \\
\textsf{IslandCountry} & \equiv & \textsf{Country} \sqcap \neg \exists \textsf{hasTerritoryIn}.\textsf{Continent} \\
\textsf{MediterrenaenCountry} & \equiv & \textsf{Country} \sqcap \exists \textsf{hasBorderTo}.\{MediterrenaenSea\} \\
\end{array}
\end{displaymath}
Moreover, let our ABox $\A_{countries}$ contain the individuals \emph{Syria, 
Turkey, France, Germany, Switzerland, USA} and 
assume we are interested in the subsumption relationships between the concept names 
\textsf{AsianCountry}, \textsf{EUmember}, \textsf{EuropeanCountry}, \textsf{G8member}
and \textsf{MediterreneanCountry}. Table \ref{tab:beforeCompletion} shows the 
partial context induced by $\A_{countries}$, and
Table \ref{tab:completionSteps} shows the questions asked by the completion
algorithm and the answers given to these questions. In order to save space,
the names of the concepts are shortened in both tables.
The questions with positive answers result in extension of the TBox with the 
following GCIs:
\begin{displaymath}
\begin{array}{rcl}
\textsf{G8member} \sqcap \textsf{MediterraneanCountry}  & \sqsubseteq & 
  \textsf{EUmember} \sqcap \textsf{EuropeanCountry} \\
\textsf{EUmember} \sqcap \textsf{G8member} & \sqsubseteq & \textsf{EuropeanCountry} \\
\textsf{AsianCountry} \sqcap \textsf{EUmember} & \sqsubseteq & 
  \textsf{MediterraneanCountry} \\
\textsf{AsianCountry} \sqcap \textsf{EUmember} \ \sqcap & & \\
  \textsf{EuropeanCountry} \sqcap \textsf{MediterraneanCountry} & \sqsubseteq & 
  \textsf{G8member} \\
\end{array}
\end{displaymath}
\begin{table}[t]
\begin{center}
\begin{tabular}{|c|c|c|c|c|c|}
\hline
$\A_{countries}$ & Asian& EU& European& G8& Mediterranean\\
\hline \hline
Syria & + & - & - & - & + \\
\hline
Turkey & + & - & + & - & + \\
\hline
France & - & + & + & + & + \\
\hline
Germany & - & + & + & + & - \\
\hline
Switzerland & - & - & + & - & - \\
\hline
USA & - & - & - & + & - \\
\hline
\end{tabular}
\end{center}
\caption{The partial context before completion}
\label{tab:beforeCompletion}
\end{table}

\begin{table}[h]
\begin{center}
\begin{tabular}{|c|c|c|}
\hline
Question & Answer & Counterex. \\
\hline \hline
\{G8, Mediterranean\}  $\rightarrow$ \{EU, European\}? & yes & - \\
\hline
\{European, G8\} $\rightarrow$ \{EU\}? & no & Russia \\
\hline
\{EU\} $\rightarrow$ \{European, G8\}? & no & Cyprus \\
\hline
\{EU, G8\} $\rightarrow$ \{European\}? & yes & - \\
\hline
\{EU, European\} $\rightarrow$ \{G8\}? & no & Spain \\
\hline
\{Asian, G8\} $\rightarrow$ \{European\}? & no & Japan \\
\hline
\{Asian, EU\} $\rightarrow$ \{Mediterranean\}? & yes & - \\
\hline
\{Asian, EU, European, Mediterranean\} $\rightarrow$ \{G8\}? & yes & - \\
\hline
\end{tabular}
\end{center}
\caption{Execution of the ontology completion algorithm
$(\T_{countries},\A_{countries})$}
\label{tab:completionSteps}
\end{table}
Moreover, the questions with negative answers result in extension of the ABox with
the individuals \emph{Russia, Cyprus, Spain} and \emph{Japan}. The partial context
induced by the resulting ABox $\A_{countries}'$ is shown in Table 
\ref{tab:afterCompletion}. The resulting knowledge base $(\T_{countries}', \A_{countries}')$ 
is complete w.r.t. the initially selected concept names.

\begin{table}[h]
\begin{center}
\begin{tabular}{|c|c|c|c|c|c|}
\hline
$\A_{countries}'$ & Asian& EU& European& G8& Mediterranean\\
\hline \hline
Syria & + & - & - & - & + \\
\hline
Turkey & + & - & + & - & + \\
\hline
France & - & + & + & + & + \\
\hline
Germany & - & + & + & + & - \\
\hline
Switzerland & - & - & + & - & - \\
\hline
USA & - & - & - & + & - \\
\hline
Russia & + & - & + & + & - \\
\hline
Cyprus & + & + & - & - & + \\
\hline
Spain & - & + & + & - & + \\
\hline
Japan & + & - & - & + & - \\
\hline
\end{tabular}
\end{center}
\caption{The partial context after completion}
\label{tab:afterCompletion}
\end{table}
\hfill $\diamond$
\end{example}

Based on on the described approach, we implemented a first experimental 
version of a DL knowledge base completion tool as an extension for the
Swoop ontology editor using Pellet as the underlying
reasoner. 
A first evaluation of this tool on the OWL ontology for human protein 
phosphatases with biologists as experts, was quite promising, but also showed
that the tool must be improved in order to be useful in practice. In particular,
we have observed that the experts sometimes make errors when answering queries.
Thus, the tool should support the expert in detecting such errors, and also make
it possible to correct errors without having to restart the completion process
from scratch. Another usability issue on the wish list of our experts was to 
allow the postponement of answering certain questions, while continuing the 
completion process with other questions.

In a follow-up paper~\cite{BaSe09} we have addressed these usability issues. 
We have improved the method in such a way that at any time during completion
the expert can pause the process, see all of her previous answers or 
changes to the knowledge base, 'undo' some of those changes, and continue
completion. Here we of course paid attention that the expert does not have to
answer the same questions she has answered before pausing the process. We have
achieved this by saving previous answers, and using them as background 
knowledge when the expert continues completion. The other wish of our experts,
namely postponing questions was solved pausing completion, changing the
order of attributes, and restarting the completion with previous answers 
as background knowledge. In theory, this method might not postpone a
question, thus the expert might be asked the last question again. However,
in practice the method turned out to be useful in many cases when the expert
was not able to answer a particular question and wanted to get another one.
We have implemented our ontology completion method together with these
usability issues as a plugin for the Prot\'eg\'e ontology editor under the 
name \textsc{OntoComP}.\footnote{\url{http://ontocomp.googlecode.com}}

\section{Conclusion}
\label{sect:conclusion}

We have summarized the work done in combining DLs and FCA. The research
done in this field mainly falls under two categories:
 1) efforts to enrich the language of FCA by borrowing constructors from
 DL languages, and  2) efforts to employ FCA methods in the solution of 
 problems encountered in knowledge representation with DLs. For each of these
categories we have given pointers and shortly described the relevant work
in the literature. We have also described our own contributions, which are
mainly under the second category. 

Recent developments in information technologies like social networks, Web 2.0
applications and semantic web applications 
are bringing up new challenges for representing vast amounts of knowledge and 
analyzing huge amounts of data rapidly generated by these applications.
The two research areas we have discussed here, namely DLs and FCA, are lying
at the core of representing knowledge, and analyzing data, respectively. 
We are confident that these new challenges will enable new fruitful 
cooperations between these two research fields.
\bibliography{/home/baris/paper/bib/fca,/home/baris/paper/bib/dl,/home/baris/paper/bib/misc,/home/baris/paper/bib/db,/home/baris/paper/bib/cc,/home/baris/paper/bib/hypergraphs}
\bibliographystyle{abbrv}

\end{document}